\documentclass[fleqn,usenatbib]{mnras}
\usepackage{hyperref}
\usepackage{newtxtext,newtxmath,times,verbatim}

\usepackage[T1]{fontenc}
\usepackage{ae,aecompl}

\usepackage{graphicx}	
\usepackage{amsmath}	
\usepackage{amssymb}	
\DeclareGraphicsExtensions{.png,.pdf}

\author[Smith \& Geach]{Michael~J.~Smith\thanks{mikejamesjsmith@gmail.com}  \&
James E. Geach\thanks{j.geach@herts.ac.uk }
\\
Centre for Astrophysics Research, School of Physics, Astronomy \& Mathematics, University of Hertfordshire, Hatfield, AL10 9AB\\
Centre of Data Innovation Research, School of Physics, Astronomy \& Mathematics, University of Hertfordshire, Hatfield, AL10 9AB}

\pubyear{2019}

\title[Generative deep fields]{Generative deep fields: arbitrarily sized, random synthetic astronomical images through deep learning}

\begin{document}
\label{firstpage}
\pagerange{\pageref{firstpage}--\pageref{lastpage}}
\maketitle

\begin{abstract}
Generative Adversarial Networks (GANs) are a class of artificial
neural network that can produce realistic, but artificial, images that resemble those in a training set.
In typical GAN architectures these images are small, but a variant known as Spatial--GANs (SGANs) can generate arbitrarily large images, provided training images exhibit some level of periodicity.
Deep extragalactic imaging surveys meet this criteria due to the cosmological tenet of isotropy.
Here we train an SGAN to generate images resembling the iconic \emph{Hubble Space Telescope} eXtreme Deep Field (XDF).
We show that the properties of `galaxies' in generated images have a high level of fidelity with galaxies in the real XDF in terms of abundance, morphology, magnitude distributions and colours. As a demonstration we have generated a 7.6-billion pixel `generative deep field' spanning 1.45\,degrees.
The technique can be generalised to any appropriate imaging training set, offering a new purely data-driven approach for producing realistic mock surveys and synthetic data at scale, in astrophysics and beyond. 
\end{abstract}


\section{Introduction}

Synthetic, or mock, data plays an important role in the interpretation of observations, as it provides a means to test a theoretical framework, as a tool to explore biases or systematics in data analysis, or to help design future experiments. In astrophysics, like many fields, a standard route to generating synthetic data is to use an analytic or semi-analytic model, or numerical simulation, to generate synthetic data that mimic the observations  \citep[e.g.,][]{cole98,Obreschkow09,mandelbaum}. The most common form of observation in astrophysics is digital imaging, and deep extragalactic imaging surveys have transformed our understanding of the Universe \citep{hdf,scoville07}. 

Current methods to generate synthetic deep fields include the projection of volume- and resolution-limited hydrodynamical cosmological simulations \citep[e.g.,][]{eagle,vogelsberger} into lightcones \citep{illustrus}, requiring a treatment for the transport of radiation through the volume and modelling of a particular instrument response \citep{sunrise,trayford}. Alternatively, mock deep fields can be created by taking an input catalogue containing the positions and properties of fake galaxies and applying models to describe their light profiles \citep{skymaker,dobke,galsim}. However, with a large, representative data set, an alternative approach is to use empirical data itself to construct new, realistic, but synthetic observations.

\begin{figure*}
    \centering{\includegraphics[width=\textwidth]{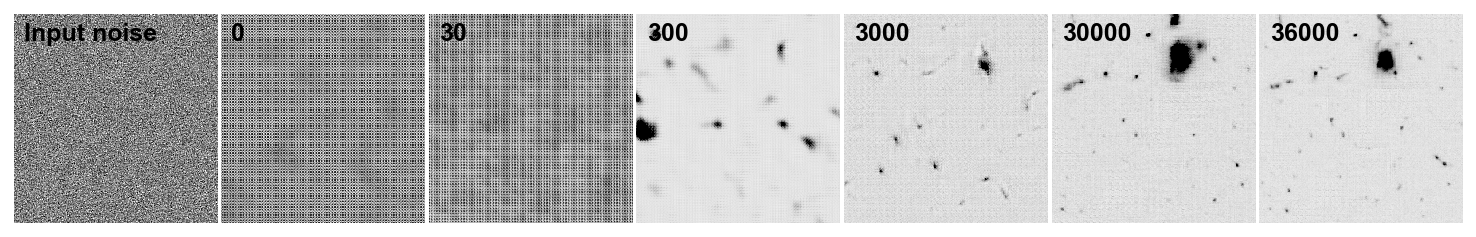}}
    \caption{Evolution of SGAN training. The images show a 256$\times$256 pixel generated image for the F775W channel after 0, 30, 300, 3000, 30,000 and 36,000 epochs using the same input latent noise vector, $z$. The distribution of input (Gaussian) noise before training is shown in the first panel. Each image is linearly scaled with a 0.5--99.5\% percentile clip. Structure becomes apparent after a few hundred training epochs and is continuously refined as training progresses. The total training time to 36,000 epochs is 1,280 hours on an NVIDIA Tesla K40c GPU.}
\end{figure*}

Generative Adversarial Networks (GANs) are a type of deep learning algorithm that can generate new samples from a probability distribution learnt from a representative training set \citep{goodfellow2014}.
The adversarial aspect of the algorithm refers to the use of two neural networks -- a generator, ${G}$ and discriminator, ${D}$ -- that compete during training.
${G}$ tries to estimate the probability distribution of the input data by producing samples that aim to trick ${D}$, which is estimating the probability that the generated sample came from the training set.
Training is a `minimax' game where ${G}$ is trying to maximize the likelihood that ${D}$ predicts the generated samples are from the real data.
The generator transforms a `latent' vector $z$, into an output $G(z)$.
Meanwhile, the discriminator takes either the output of the generator $G(z)$, or a real data example $x$, and transforms this input into an output, $D(G(z))$ or $D(x)$.
The output can be thought of as the probability that $G(z)$ is indistinguishable from $x$. The networks are trained through gradient descent \citep{robbins1951} until $G(z)$'s distribution closely matches the distribution of $x$.
After training, the generator can produce convincing images resembling those in the training set. These generated images can be thought of as random draws from the probability distribution estimated by the generator that describes the distribution from which pixels in the training image were sampled.

In this work we present a method exploiting GANs to generate realistic, but random, extragalactic deep field images of arbitrary size. In Section 2 we describe the `Spatial GAN' variant, and explain how it is trained to generate fake images. In Section 3 we describe our results, comparing `galaxies' detected in the fake images to galaxies detected in the training image. We discuss and summarise the results and highlight limitations and scope for improvement and future work in Section 4. Magnitudes are all quoted on the AB system.

\section{Method}

\subsection{Spatial--GANs}

A Spatial--GAN \cite[SGAN,][]{jetchev2016} is a fully convolutional GAN that uses variably sized 2D latent vector arrays as the generator input $z$.
This is in contrast to a standard GAN, which uses a 1D latent vector.
In a standard GAN a dense neural layer is used to connect and reshape the latent vector so that a convolutional layer can operate on it.
This lack of a fully-convolutional architecture means that a standard GAN can only produce images of a single, fixed shape.
An SGAN replaces all dense neural layers in both its generator and its discriminator with convolutional layers.
This allows input of variably sized image--latent vector array pairs.
In an SGAN the latent vector arrays are upsampled using deconvolving layers in the generator, and both generated and real images are downsampled in the discriminator via convolving layers.
Since $z$ can be varied, an SGAN can be used to create an image of any size, even one much larger than seen in the training set.
If the training images exhibit periodicity, the SGAN's generator will also learn to exhibit the same periodicity \citep{jetchev2016}.

\subsection{Training set}

We use a training set comprised of the F814W, F775W and F660W bands of the {\it Hubble Space Telescope} eXtreme Deep Field \cite[XDF,][]{XDF}, with all images aligned and sampled on the same 60-mas grid\footnote{\href{https://archive.stsci.edu/prepds/xdf}{https://archive.stsci.edu/prepds/xdf}}.
The only other preprocessing is a channel-wise clip at the 99.99th percentile.
Unlike many GAN approaches that use training images scaled to an 8-bit depth per channel, we train using the full floating point dynamic range of the data, with 32-bit depth per channel.
Training images are sampled from the full XDF image by cropping the image at random positions with crop sizes of 64, 128 or 256 pixels, with the size fixed for each batch.
Corresponding noise arrays of sizes 4, 8, or 16 pixels, are sampled from a Gaussian distribution with a zero mean and unity variance, and passed to the generator.
Each noise array has a channel axis of size 50.

\begin{figure*}
    \centering{\includegraphics[width=0.5\textwidth]{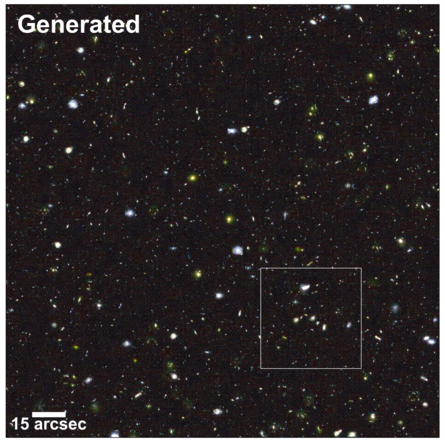}\includegraphics[width=0.5\textwidth]{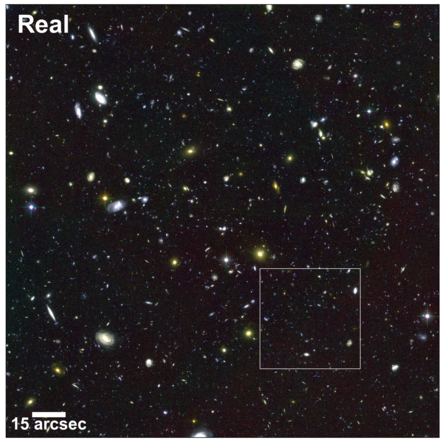}}
    \centering{\includegraphics[width=0.5\textwidth]{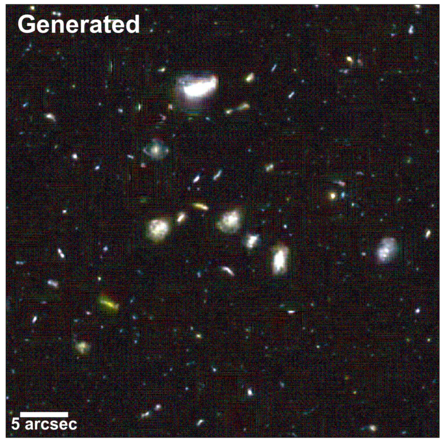}\includegraphics[width=0.5\textwidth]{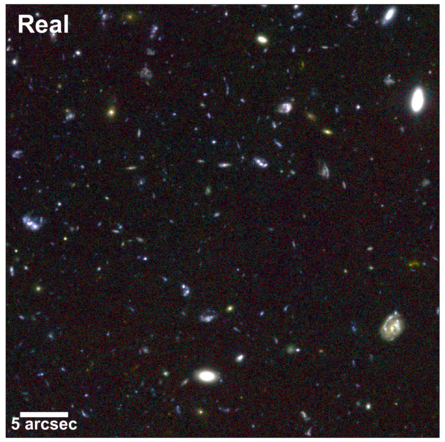}}
    \caption{Example of a generated deep field. The top left panel shows a generated deep field with the RGB channels corresponding to the F814W, F775W and F606W bands. The size of the generated image was set to 3.2$'$, spanning the full XDF (top right), which has a 60\,mas pixel$^{-1}$ scale. The lower panels show 45$''$ zoom-ins of the generated and real fields.}
\end{figure*}

\subsection{Architecture and training}

The generator is comprised of one initial deconvolutional layer with a stride of 2 and a kernel size of 4.
Three deconvolutional sets are then applied, each comprising of one layer with a stride of 2, followed by three layers with strides of 1.
Each layer in these sets have kernel sizes of 4.
These layers have an exponential linear unit (ELU) activation \citep{clevert2015}.
The generator's output layer is also deconvolutional, with a stride of 1, a kernel size of 3, and 3 filters.
The output layer has a sigmoid activation function, \begin{equation}
    S(x)=(1+e^{-x})^{-1}.
\end{equation}
The discriminator comprises of 5 initial convolutional layers, each with a stride of 2, and a kernel size of 4.
All initial layers have a Leaky Rectified Linear Unit (ReLU) activation \citep{maas2013}.
2D global average pooling is applied to the final convolutional layer, and a dense layer is used to connect the global average pool to a binary classification output.

The discriminator uses a relativistic average loss \citep{AJM2018}.
A relativistic discriminator estimates whether the incoming data are more realistic than a random sample of the opposing type.
To understand why this is important, consider that in a standard GAN a perfect generator will cause the discriminator to define all incoming data as `real'.
However, it is known that exactly half of an incoming batch is real data.
Therefore, if the generator is producing flawless fakes, the discriminator should assume that each sample has a 50\% probability of being real \citep{AJM2018}.
To take this prior knowledge into account, the output function of a non-relativistic discriminator is modified from $D(x) = S(C(x))$, where $C(x)$ is the output of the final layer with no applied activation function, to ${D}_R(x)$

\begin{equation}
  D_R(x) =
  \begin{cases}
    S(C(x) -  \mathbb{E}_{x_f \sim  \mathbb{Q}}C(x_f))\ \text{for real $x$},\\
    S(C(x) -  \mathbb{E}_{x_r \sim  \mathbb{P}}C(x_r))\ \text{for generated $x$},
  \end{cases}
\end{equation}
where $S$ is the sigmoid activation function.
The second parts of the real and generated $D_R(x)$ are effectively the average discriminator value for fake images, $x_f$, and real images, $x_r$, respectively.
The use of a relativistic discriminator leads to stable training on a difficult data set of large images, where a standard discriminator fails \citep{AJM2018}.
The discriminator is packed to stabilize training \citep{lin2017}.
To pack the discriminator, two images from the same class are concatenated along their channel axes and fed into the discriminator as a single sample.
Packing the discriminator in this way reduces the possibility of mode collapse \citep{lin2017}.
The Adam optimizer \citep{kingma2014} is used, with a learning rate of 0.0002. The learning rate was determined through a manual search as the maximum rate that yields stable training.

The SGAN is trained on an NVIDIA Tesla K40c GPU for 36,000 epochs of 30 batches with a batch size of 128. Each epoch requires approximately 120 seconds, depending on the batch crop sizes. The evolution of the generated images for a fixed latent noise vector $z$ is shown in Figure~1, showing the emergence of amorphous structure and then refinement into structures resembling galaxies with increasing training time.

\begin{figure}
    \centerline{\includegraphics[width=0.475\textwidth]{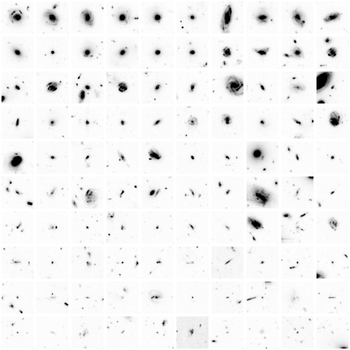}}
    \caption{Thumbnail images of real and generated `galaxies'. Each image is 7.2$''$ across and shows the F775W channel on an identical linear grayscale for 100 sources selected in the range 21--26\,mag. Fifty targets are selected from each of the real and generated catalogues and displayed in random order in row-wise ascending order of magnitude.}
\end{figure}

\begin{figure*}
    \centering{\includegraphics[width=0.33\textwidth]{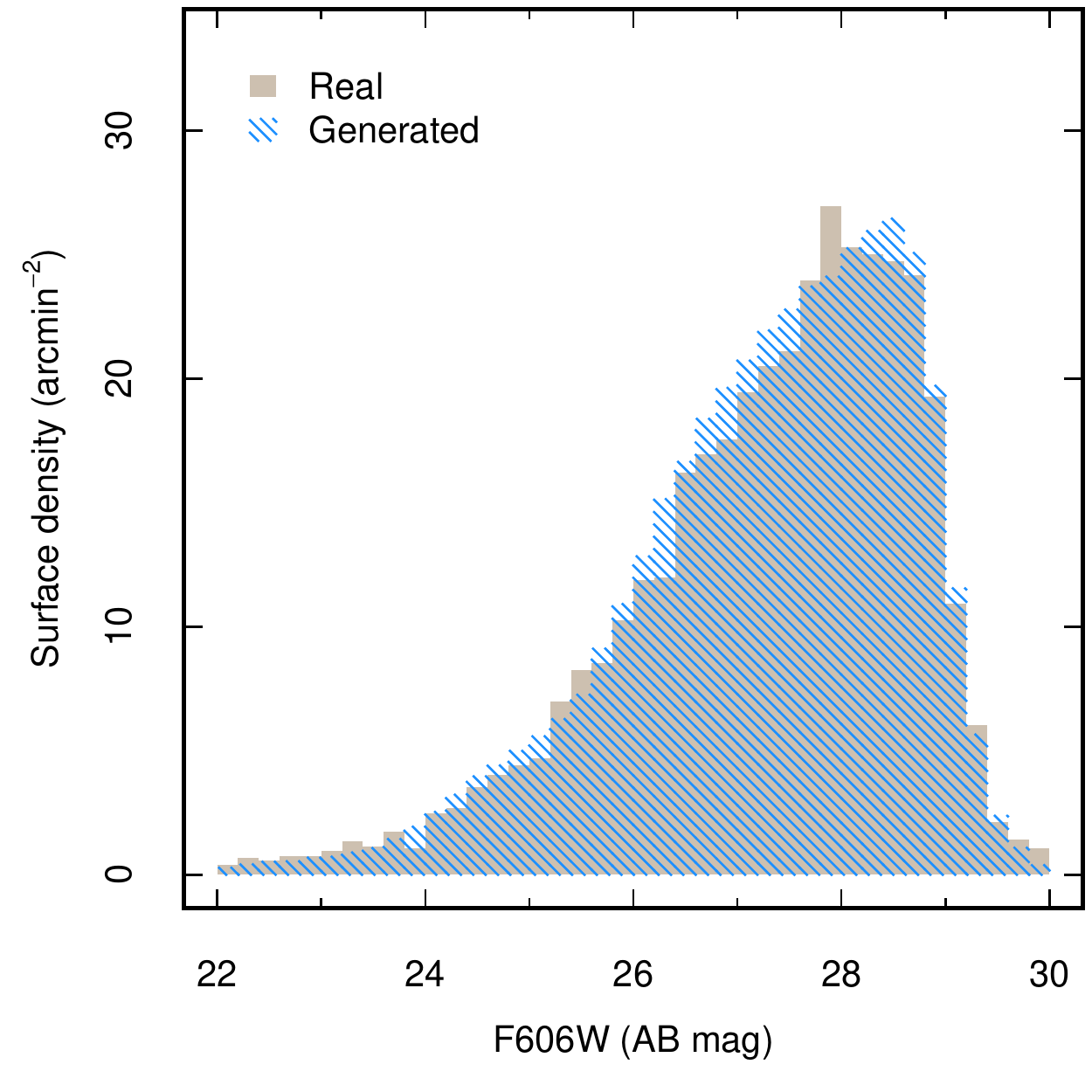}\includegraphics[width=0.33\textwidth]{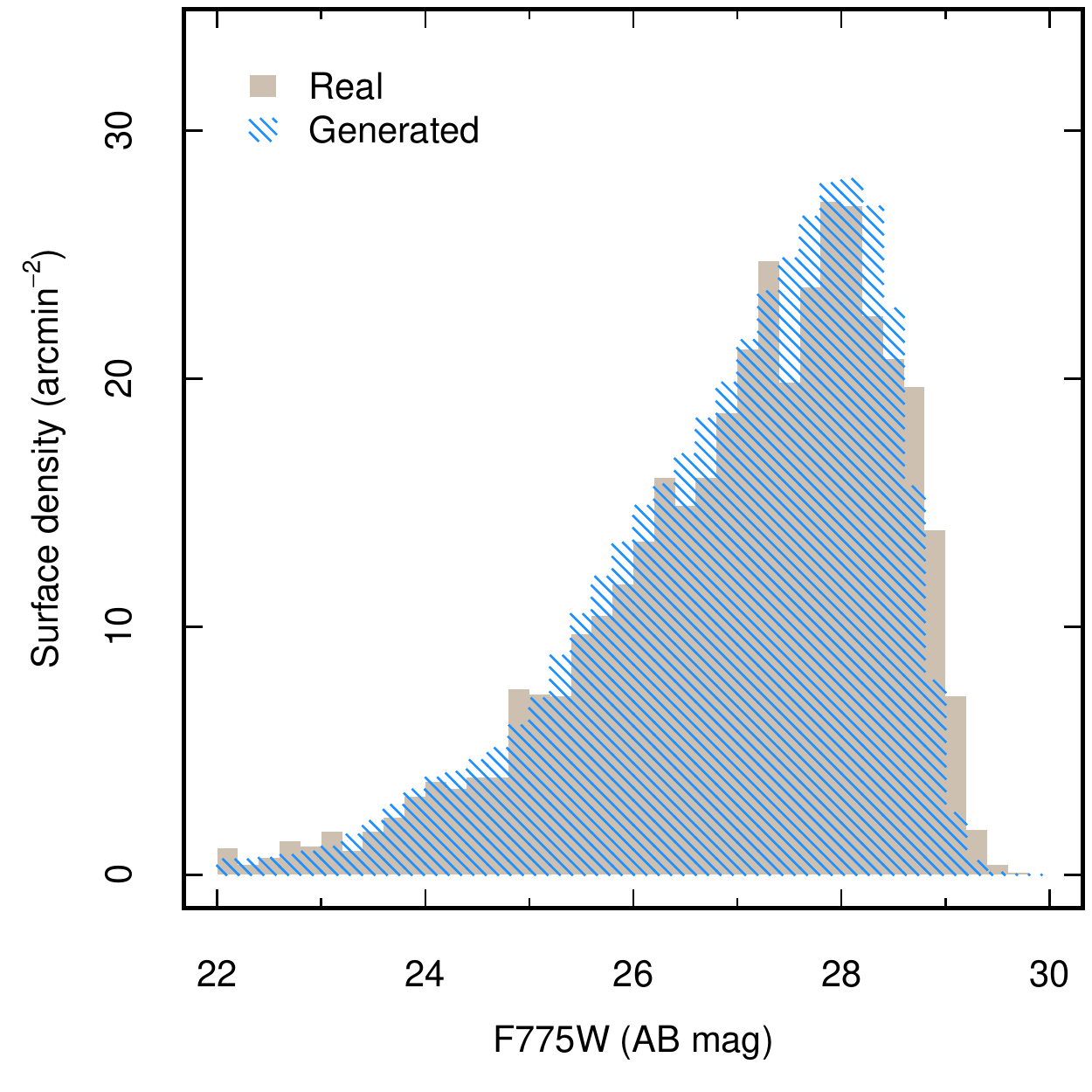}\includegraphics[width=0.33\textwidth]{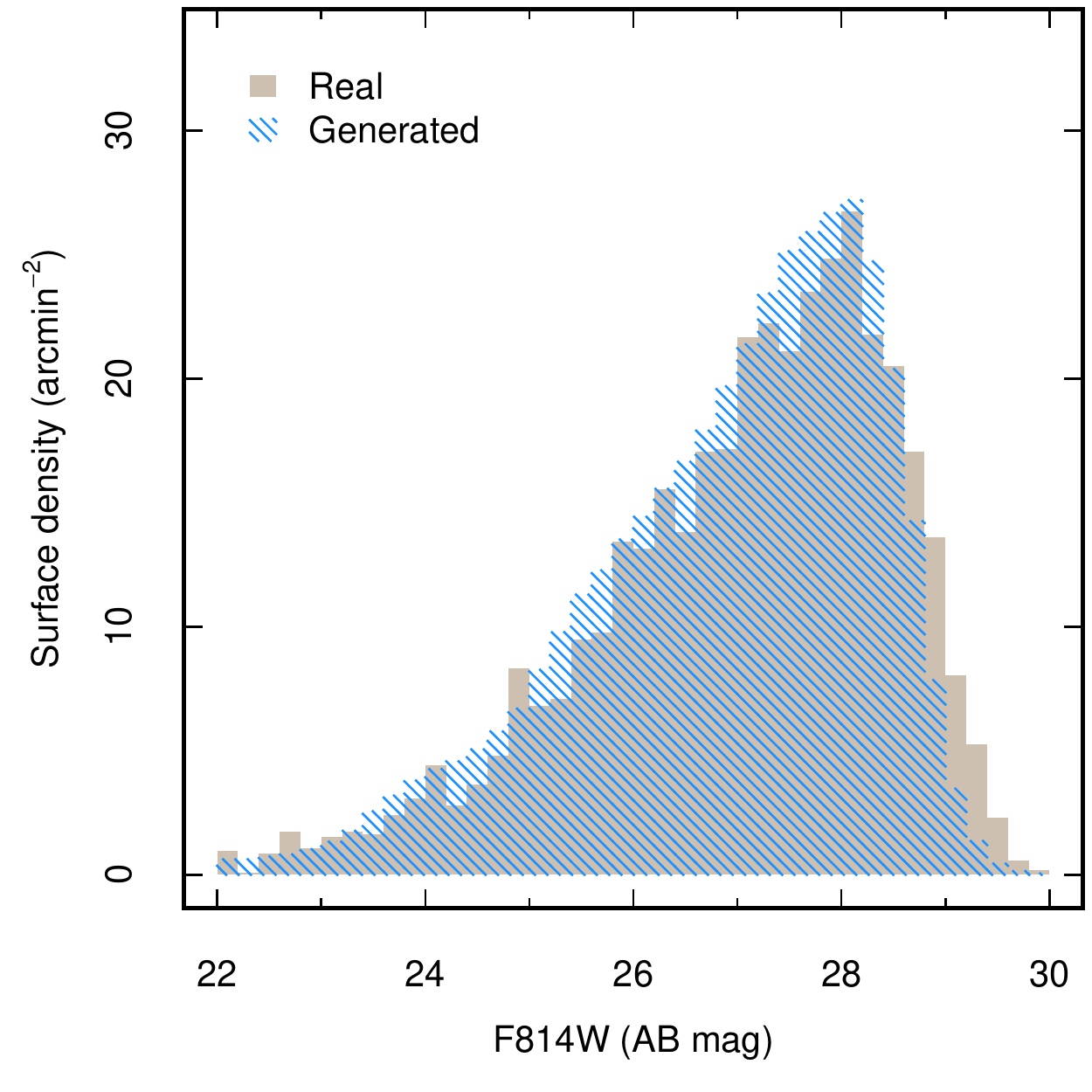}}
\caption{The photometric properties of galaxies detected in the real XDF and ensemble of generated images. The histograms show the magnitude distributions of the galaxies detected in the band-merged  image and measured in each of the F606W, F775W and F814W bands. The photometric zeropoints applied to the generated data is identical to the zeropoints calibrated for the real data.}
\end{figure*}

\section{Results}

A self-ensemble of 21 GANs is created by taking every tenth epoch's model from a range of 200 epochs. A Convolutional Neural Network (CNN) ensemble is a combination of different CNNs trained from different weight initialisations on the same data. A self-ensemble is an ensemble of CNNs trained from the same initialisation, but taken at different points in the training cycle. Ensembles of CNNs can produce significantly more accurate predictions when compared to a single CNN \citep{Nanni2018, Paul2018}. This increase in accuracy has been shown by \cite{wang2016} and \cite{mordido2018} to also be present in GAN ensembles.

Four outputs were taken from each of these 21 GANs, resulting in 84 total XDF simulations. To generate a simulation set, each of the trained generators is fed a $z$ vector that has shape $[b, 202, 202, 50]$, where $b = 4$ is the batch size.
The generators up-sample $z$ to a shape $[b, 3232, 3232, 3]$, matching the shape of the training image.
The generation time for single realization of the XDF is 15\,seconds for all three bands on an Intel Xeon CPU~E5-2650~v3.
In principle, the generated image can be of arbitrary contiguous size by using seamless tessellation.
Seamless tessellation can be achieved with SGAN by having adjacent tiles $a$ and $b$ share a portion of boundary $z$ values.
This creates a shared area where $G(z_a)$ and $G(z_b)$ have an equal output.
A seamless tile is made by cropping at the midpoint of the shared noise and concatenating $G(z_a)$ and $G(z_b)$. To illustrate, we have generated a contiguous 87040$\times$87040 pixel ($1.45^\circ\times1.45^\circ$) version of the XDF that can be examined online at \href{http://star.herts.ac.uk/~jgeach/gdf.html}{http://star.herts.ac.uk/$\sim$jgeach/gdf.html}.

Figure~2 presents an example of a generated field in comparison to the real XDF, combining the F814W, F775W and F606W bands into an RGB colour composite.
We show the full field, spanning approximately 3$'$, and a zoom-in of a 45$''$ region for closer inspection.
The generated image is reasonably convincing as an extragalactic deep field from the point of view of cursory visual inspection, although like many GAN-generated images, on closer inspection there are artifacts (such as low surface brightness features, pixelization, and discontinuities not seen in the real image) that betray the counterfeit.
The GAN also cannot reproduce some of the larger galaxies with the same level of detail as the real image.
Nevertheless, the generated image does contains a mixture of `early' and `late' type galaxies amongst a more numerous field of background sources.
The early and late types have the correct color and morphology, i.e.\ classic red/elliptical and blue/spiral respectively, and background sources have the clumpy and disturbed morphology typically seen in the high redshift galaxy population.
Figure~3 presents single-band (F775W) thumbnail images of a sample of 100 randomly chosen sources with F775W AB magnitudes in the range 21--26\,mag.
50 sources are selected from the real image and 50 are selected from the generated images, but the thumbnails have been randomly shuffled in the Figure to demonstrate to the reader the visual similarity of individual galaxies in the generated and real data.

To test whether the similarity between the generated and real image is more than superficial, we extract `galaxies' from the generated images and compare to those in the real XDF.
We use the source finder {\it SExtractor} \citep{sextractor} to detect galaxies in the combined F606W+F775W+F814W (pixelwise-mean) images, measuring the corresponding source flux in the individual bands.
Sources are identified with the criteria that 5 contiguous pixels have a signal $\geq$5$\sigma$ above the local background.
The same detection procedure is applied to the real and generated images, and the photometric zeropoints for the generated images are identical to the real data.
Figure~4 compares the magnitude distribution of sources detected in the real field and ensemble of generated fields measured in each of the F606W, F775W and F814W bands. The generated and real distributions bear close statistical resemblance. The absolute difference in the generated and real median magnitudes in each of the F606W, F775W and F814W bands is within 0.02\,magnitudes, and a the generated images produce the correct abundance of galaxies across the full magnitude range in every band.
The {\it p}-values given by a two-sided Kolmogorov-Smirnov test for a magnitude limit of 28\,mag (approximately the completeness limit of the catalogue for our detection criteria) are  0.16, 0.63 and 0.75, indicating that we cannot say with confidence that the distributions are drawn from different parent populations, and are therefore  statistically similar.

\section{Discussion and Conclusions}

While the SGAN can produce arbitrarily-sized simulations of the XDF, there are clear limitations of this technique.
The primary limitation is that the generated images are of course totally biased to the training set.
The consequence is that a generated image much larger than the training data will not be cosmologically representative.
This is simply because the XDF probes a volume too small to contain rare objects such as, for example, clusters of galaxies and the generator cannot produce examples of objects it has not seen.
This problem is simple to alleviate by using a much larger and representative training set.
In the real Universe, the positions of galaxies are correlated, even on large angular scales, due to the presence of cosmological large-scale structure.

Another limitation is that this clustering information is not present in the generated images, although correlations on the scales of the crop size will be somewhat preserved.
Finally, we were not able to achieve stable training with more than three photometric bands.
This issue could be resolved with more regularization: adding dropout \citep{srivastava2014}, or spectral normalisation \citep{miyato2018} could allow one to generate additional bands, although at the cost of a longer training time, and a larger memory footprint during training.

An advantage of this technique is that it is empirically driven, since the data itself is used as the model.
By training directly on imaging data the SGAN simultaneously encodes information about the instrumental response as well as the `galaxy' population, thus circumventing the need to make modelling assumptions about either, as is the case in other approaches.
The corollary to this is that in the current approach we cannot disentangle the instrument response from the underlying galaxy population, although one could envision approaches to tackle this challenge.
For example, a method similar to that used by \cite{schawinski2017}, where a GAN is used to remove noise in astronomical imagery, could be employed.
Specifically, a GAN with a UNet-like \citep{ronneberger2015} generator would learn a transformation between an image with an entangled instrument response, and one without. 

Recent advances in GAN training techniques have resulted in impressively high fidelity image output \citep{karras2017,brock2018,zhang2018}.
Similar methods could be implemented to improve the quality of the generated deep fields.
Training on a larger set of imaging data, with a increased batch size \citep{brock2018}, may also produce more representative simulations on a deeper network.
The addition of a GAN architecture that allows for control over the output, such as InfoGAN \citep{chen2016} or Conditional GAN \citep{mirza2014} could also be useful when creating mock surveys because they would allow control over the frequency of particular objects of interest, or over the makeup of background noise.

This method could be used to generate at scale entirely artificial, but realistic, image realizations for the design, development, and exploitation of new surveys.
For example, one could assemble large training sets for instance segmentation and classification of galaxies.
In the early stages of a new survey, relatively small amounts of data could be collected, but then expanded to a level useful for training deep learning models using the generative method described here.
Segmentation and classification algorithms could then be trained on the generated data, and then applied to new data, allowing far faster deployment of astronomical deep learning algorithms than would otherwise be possible, potentially accelerating the exploitation of new survey data.

\section*{Acknowledgements}

M.J.S. acknowledges the hospitality of Queen's University during much of this work. J.E.G. is supported by the Royal Society. We gratefully acknowledge the support of NVIDIA Corporation with the donation of the Tesla K40 GPU used for this research, which also made use of the University of Hertfordshire
high-performance computing facility (\href{http://uhhpc.herts.ac.uk}{http://uhhpc.herts.ac.uk}). The SGAN code can be obtained at \href{https://github.com/Smith42/XDF-GAN}{https://github.com/Smith42/XDF-GAN}. The 7.6-billion pixel version of the generated XDF can be viewed at \href{http://star.herts.ac.uk/~jgeach/gdf.html}{http://star.herts.ac.uk/$\sim$jgeach/gdf.html}.

\bibliographystyle{mnras}
\bibliography{gdf_v1.bib}

\end{document}